\documentclass[12pt,preprint]{aastex}

\slugcomment{accepted to The Astrophysical Journal Letters}

\begin{document}

\title{WHAM Observations of H$\alpha$ from High Velocity Clouds: Are They
Galactic or Extragalactic?}

\author{S. L. Tufte and J. D. Wilson}
\affil{Department of Physics, Lewis \& Clark College, 0615 SW Palatine
Hill Road, Portland, OR, 97219; tufte@lclark.edu, jdwilson@lclark.edu}

\and

\author{G. J. Madsen, L. M. Haffner, \& R. J. Reynolds}
\affil{Astronomy Department, University of Wisconsin - Madison, 475
N. Charter Street, Madison, WI, 53706; madsen@astro.wisc.edu,
haffner@astro.wisc.edu, reynolds@astro.wisc.edu}

\pagebreak

\begin{abstract}
It has been suggested that high velocity clouds may be distributed
throughout the Local Group and are therefore not in general associated
with the Milky Way galaxy.  With the aim of testing this hypothesis,
we have made observations in the H$\alpha$ line of high velocity
clouds selected as the most likely candidates for being at larger than
average distances.  We have found H$\alpha$ emission from 4 out of 5
of the observed clouds, suggesting that the clouds under study are
being illuminated by a Lyman continuum flux greater than that of the
metagalactic ionizing radiation.  Therefore, it appears likely that
these clouds are in the Galactic halo and not distributed throughout
the Local Group.
\end{abstract}


\section{Introduction}

Our understanding of the nature of high-velocity clouds (HVCs),
defined as interstellar clouds moving at velocities not compatible
with a simple model of differential Galactic rotation, is severely
limited by our lack of knowledge of their distances.  Except in a few
isolated cases (see Wakker 2001), their distances are very poorly
constrained.  Several authors (Blitz et al. 1999; Braun \& Burton
1999) have suggested that a subclass of the HVCs are dispersed
throughout the Local group of galaxies, the remnants of its formation.
This would place them much farther away (100 kpc to 1000 kpc) than
other models that place them in the Galactic halo at $\lesssim$ 10 kpc
distances (e.g. Oort 1970, Bregman 1980).  Given their angular size
and neutral column density, the larger distances would make them very
massive objects.

We have tested this hypothesis by measuring the H$\alpha$ intensity
toward a collection of HVCs whose properties open the possibility that
they may be at greater than average distances.  If it is indeed true
that these clouds lie at great distances from the Galaxy, their
neutral gas should not be substantially ionized by the weak
metagalactic ionizing flux.  Weymann et al. (2001) set a 2$\sigma$
upper limit of 8 mR for the H$\alpha$ intensity toward an
intergalactic H~I cloud and inferred upper limits to the metagalactic
ionizing flux, $\Phi_{\rm o}$, between 2.5 $\times$ 10$^{3}$ cm$^{-2}$
s$^{-1}$ and 1.0 $\times$ 10$^{4}$ cm$^{-2}$ s$^{-1}$ with a preferred
value of 5 $\times$ 10$^{3}$ cm$^{-2}$ s$^{-1}$, the range resulting
from uncertainties in the cloud geometry (see also Shull et al. 1999
for a review of constraints on $\Phi_{\rm o}$).  However, if the cloud
is in the vicinity of the Milky Way, a source of Lyman continuum
photons, it will be more strongly ionized and therefore glowing in
H$\alpha$ (see Bland-Hawthorn \& Maloney 1999).  In this case the
observations are a measure of the escape fraction of ionizing photons
from the Milky Way, which is important both in understanding the
radiation transfer within the Galaxy and for understanding the impact
of the Galaxy on its environment.

\section{Observations}
We used the Wisconsin H-Alpha Mapper (WHAM) for all observations.
This dual-etalon Fabry-Perot instrument is well suited to the task of
measuring the extremely faint emission lines here investigated
(Reynolds et al. 1998, Tufte 1997.)  WHAM measures the spectrum of
emission contained within the 1\arcdeg\ observation beam with 12
km~s$^{-1}$ velocity resolution.  The HVCs were carefully selected to
be the most likely candidates for being Local Group clouds based on
their properties in the Leiden/Dwingeloo Survey (LDS) (Hartmann \&
Burton 1997).  The selection criterion included appropriate angular
size: a compact cloud is more likely to be far away due to the
correlation of angular size with distance, but if it is much smaller
than the WHAM beam, then beam dilution reduces the detection
threshold.  The clouds are isolated from any of the larger HVC
complexes or the Galactic H~I in ($l$,$b$,$v$) space.  We also
obtained guidance in selecting the observation directions from Blitz
(private communication) and from the dynamical model simulating the
formation of the Local Group presented by Blitz et al. (1999).  The
remnant fragments of H~I in this model tend to loosely correlate with
the axis formed by the Milky Way and Andromeda galaxies (see the upper
portion of Figure 13 in Blitz et al. 1999).  Also in the model, clouds
in the general longitudinal vicinity of M31 and with negative Galactic
latitudes have a strong tendency to have velocities -400 km s$^{-1}$
$<$ V$_{lsr}$ $<$ -200 km s$^{-1}$ (see their Figure 14). Our observed
clouds are clustered in this region of ($l$, $b$, $v$) space.
Independent of the model, more extreme velocities are more likely to
represent material not directly associated with our Galaxy.  As a
result of these considerations, the sightlines chosen are among the
HVCs most likely to represent ``Local Group'' clouds.

A total of 17 sightlines toward 6 ``clouds'' were observed with
exposure times per sightline ranging from 600s to 1800s.  An ``on -
off'' observing procedure was used to minimize the influence of faint
terrestrial lines near H$\alpha$ (see Tufte, Reynolds, \& Haffner 1998
for further explanation of this method).  For one of the clouds, HVC
231, we obtained inconclusive results (neither a detection nor a
convincing upper limit), and this case is not included in the results
below.

\section{Results}
Four of the five clouds were detected in H$\alpha$, and for the fifth
cloud we set a very low upper limit to the emission.  Figure 1 shows
the five H$\alpha$ sightlines selected for HVC 532 overlayed on a
contour map of the 21-cm intensity for this compact HVC.  All 21-cm
data are from the Leiden/Dwingeloo H~I survey of Hartmann \& Burton
(1997).  Figure 2 shows the H$\alpha$ and 21-cm spectrum for beam (A).
In the H$\alpha$ spectrum a clear emission feature is seen with an
intensity I$_\alpha$ = 0.14 $\pm$ 0.01 R (1 R = 10$^6$ / 4$\pi$
photons cm$^{-2}$ s$^{-1}$ sr$^{-1}$).  The H$\alpha$ feature is in
close correspondence to the 21-cm spectral feature, although it is
wider and slightly shifted to lower negative LSR velocity.  Figure 3
shows the H$\alpha$ emission from directions (B) and (D) from Figure 1
in the upper panels with the corresponding 21-cm spectra below them.
In direction (B) there is marginal evidence for H$\alpha$ and in
direction (D) there is clear evidence for H$\alpha$ emission with an
intensity of I$_\alpha$ = 0.12 $\pm$ 0.01 R, similar to direction (A).
In directions (C) and (E), no H$\alpha$ was detected, indicating that
the ionized gas does not extend much beyond the H~I for this cloud.

Table 1 summarizes the results for all of the observations included in
this work, including the HVC 532 results shown above.  The H$\alpha$
intensity toward HVC 486 is I$_{\alpha}$ = 0.13 $\pm$ 0.04 R, similar
to HVC 532.  These intensities are strikingly similar to those
measured toward the M, A, and C HVC complexes, which have intensities that
range from 0.06 R to 0.20 R (Tufte et al. 1998), and which are
believed to be in the Galactic halo (Wakker 2001).  In the direction
HVC 518 there is a clean detection of emission with I$_{\alpha}$ =
0.032 $\pm$ 0.004 R, and there is also emission from HVC 444 at a
similar level but with lower signal-to-noise.  Since the uncertainty
is usually governed by systematic effects (e.g. the inability to
perfectly subtract off terrestrial emission lines) instead of photon
statistics, the error bars were all obtained by experimenting with
different fits to the data and monitoring the residuals to determine
the range of emission parameters consistent with the data.  For the
non-detections, the results listed in Table 1 correspond to
conservative 2$\sigma$ upper limits.

In one case, HVC 394, there is no evidence of H$\alpha$ emission down
to a very low level.  Figure 4 shows the observation directions for
this cloud overlaid on 21-cm contours.  Figure 5 shows the H$\alpha$
and 21-cm spectra for direction labeled (G) in Figure 4.  There is no
sign of an emission feature at the location of the 21-cm emission.
The dotted line shows a hypothetical feature with an intensity of 0.02
R, considered here as an upper limit to the H$\alpha$ emission for
this direction.  Table 1 shows that none of the directions showed
emission, and in one case the upper limit is as low as I$_\alpha$
$\leq$ 0.01 R (2$\sigma$), one of the faintest limits yet achieved.

\section{Discussion and Conclusions}

Of the five candidate Local Group clouds for which we obtained good
measurements, there is clear H$\alpha$ emission from four.  The
intensities are around 0.1 R, typical of the intensities measured from
the large HVC complexes thought to lie in the Galactic halo.  We
conclude from this that it is unlikely that these clouds are at 100
kpc distances from our Galaxy, because the metagalactic ionizing flux
level is below that needed to produce the observed H$\alpha$ surface
brightness of even the faintest of the detections.  If they are
``Local Group'' objects, then the metagalactic ionizing flux needs to
be much higher than previously thought.  However, recent observations
of the outer H~I disk of M 31 (Madsen et al. 2001) appear to rule out
such a large ionizing flux within the Local Group.  It is more likely
that these clouds are not at great distances, but are instead in the
Galactic halo and being ionized by the Milky Way.  If this latter
scenario is true, these observations support a picture where the
distribution of ionizing flux percolating through the disk is somewhat
patchy and the Lyman continuum flux, F$_{LC}$, incident onto a cloud
from the Galactic plane averages F$_{LC}$ $\simeq$ 2 $\times$ 10$^5$
cm$^{-2}$ s$^{-1}$ (assuming 0.1 R as a representative H$\alpha$
intensity; see Tufte et al. 1998).

The extreme faintness of H$\alpha$ from HVC 394 place this measurement
among the lowest upper limits to the metagalactic radiation field,
comparable to the measurements of Madsen et al. 2001, Weymann et
al. 2001, and Vogel et al 1995.  Since the H~I cloud is optically
thick in the Lyman continuum and optically thin to H$\alpha$ photons,
each Lyman continuum photon incident on the cloud will ionize a
hydrogen atom, and each hydrogen recombination will produce on average
0.46 H$\alpha$ photons (Martin 1988; Pengally 1964; case B, T = 10$^4$
K).  If we assume that all of the H$\alpha$ arises from gas
photoionized by a uniform isotropic metagalactic ionizing flux, then
the upper limit on I$_{\alpha}$ provides direct constraints on this
quantity.  Since the WHAM beam is easily contained within the H~I
cloud (see beam F in Figure 4), the metagalactic ionizing flux
generates observable H$\alpha$ from the front and back faces of the
cloud and we approximate the geometry as a plane slab viewed normally.
Under these assumptions, the measured upper limit for HVC 394 of
I$_\alpha$ $<$ 0.01 R corresponds to an upper limit to the
metagalactic ionizing flux $\Phi_{\rm o}$ $<$ 1.1 $\times$ 10$^4$
photons cm$^{-2}$ s$^{-1}$, where $\Phi_{\rm o}$ is the incident
one-sided ionizing flux defined by Vogel et al. 1995 and Madsen et
al. 2001.  If other processes contribute to ionizing hydrogen in this
cloud, then $\Phi_{\rm o}$ must be even lower.

Perhaps HVC 518 and HVC 444 are fainter in H$\alpha$ because they are
further away than HVC 532, HVC 486, and the M, A, and C complexes.
However, with the exception of HVC 394, these results all suggest that
the clouds are located in the halo rather than the intergalactic
medium.  Corroborating evidence comes from the recent H$\alpha$
observations of Weiner et al. (2001) of a different set of compact
clouds, where they found intensities ranging from 0.04 R to 1.6 R,
with no well-measured clouds showing non-detections.  The strength of
these conclusions are currently limited by the relatively
small number of clouds sampled, and further observations of the compact
HVCs will be necessary to increase the statistics.  Also, observations
in other emission lines such as [S~II], [N~II], and [O~III] should
help to illuminate further the characteristics of the compact
population of HVCs and their relationship to the major complexes.

We are grateful to Dr. Leo Blitz for constant encouragement and
support, and for his help in selecting HVCs to observe.  S. L. T. and
J. D. W. acknowledge support from Research Corporation through a
Cottrell College Science Award.  G. J. M., L. M. H., and
R. J. R. acknowledge support from the National Science Foundation
through grant AST 96-19424.

\pagebreak

\begin{deluxetable}{lcccccccc}
\footnotesize
\tablecaption{Measured Properties of the HVC H$\alpha$ Emission}
\tablewidth{0pt}
\tablehead{
\colhead{HVC}           & \colhead{$l$}      &
\colhead{$b$}          & \colhead{V$_{H\alpha}$($\frac{km}{sec})$\tablenotemark{a}}  &
\colhead{W$_{H\alpha}$($\frac{km}{sec})$\tablenotemark{b}}     & \colhead{I$_{\alpha}$($R$)}    &
\colhead{V$_{H I}$($\frac{km}{sec})$}  & \colhead{W$_{H I}$($\frac{km}{sec})$}  &
\colhead{$N_{H}$($\frac{10^{19}}{cm^{2}}$)}
}
\startdata
532 A & 118.5 & -58.2 & -369 & 48$\pm$3 & 0.14$\pm$0.01 & -374 & 28 & 3.1 \\
532  B & 119.5 & -58.2 & -369 & 38$\pm$17 & 0.04$\pm$0.01 & -375 & 33 & 1.6   \\
532  C & 120.5 & -58.2 & -371 & \nodata &$< 0.02$ & \nodata &\nodata &\nodata \\
532 D & 117.5 & -58.2 & -362 & 30$\pm$4 & 0.12$\pm$0.01 &-371 &32& 1.7 \\
532 E & 118.5 & -59.2 & -371 &\nodata &$< 0.02$ &\nodata &\nodata &\nodata \\
486 & 158.0 &-39.0 &-290 &16$\pm$7 &0.13$\pm$0.04 &-284 & 27 & 0.5 \\
518 &104.2 &-48 &-168 &13$\pm$5 &0.032$\pm$0.004 &-170 & 25 &0.6 \\
444 & 119.2 & -30.8 & -382 & 43 $\pm$15 & 0.02$\pm$0.01 & -386 & 20 & 1.1 \\
394 F & 72 & -22 &-330 &\nodata &$< 0.01$ &-329 & 32 & 1.6 \\
394 G & 71 & -22 &-330 &\nodata &$< 0.02$ &-332 & 25 & 1.9 \\
394 H & 70 & -22 &-330 &\nodata &$< 0.02$ &-332 & 25 & 3.0 \\
394 I & 73 & -22 &-330 &\nodata &$< 0.02$ &-326 & 23 & 0.8 \\
394 J & 71 & -23 &-330 &\nodata &$< 0.02$ &-326 & 23 & 0.16 \\
\enddata

\tablenotetext{a}{The uncertainties in the velocities are $\pm$ 1 km s$^{-1}$.}
\tablenotetext{b}{The width parameters for H$\alpha$ and 21-cm components are FWHM for Gaussian fits.}

\end{deluxetable}

\clearpage

\begin{figure}[ht]
\plotone{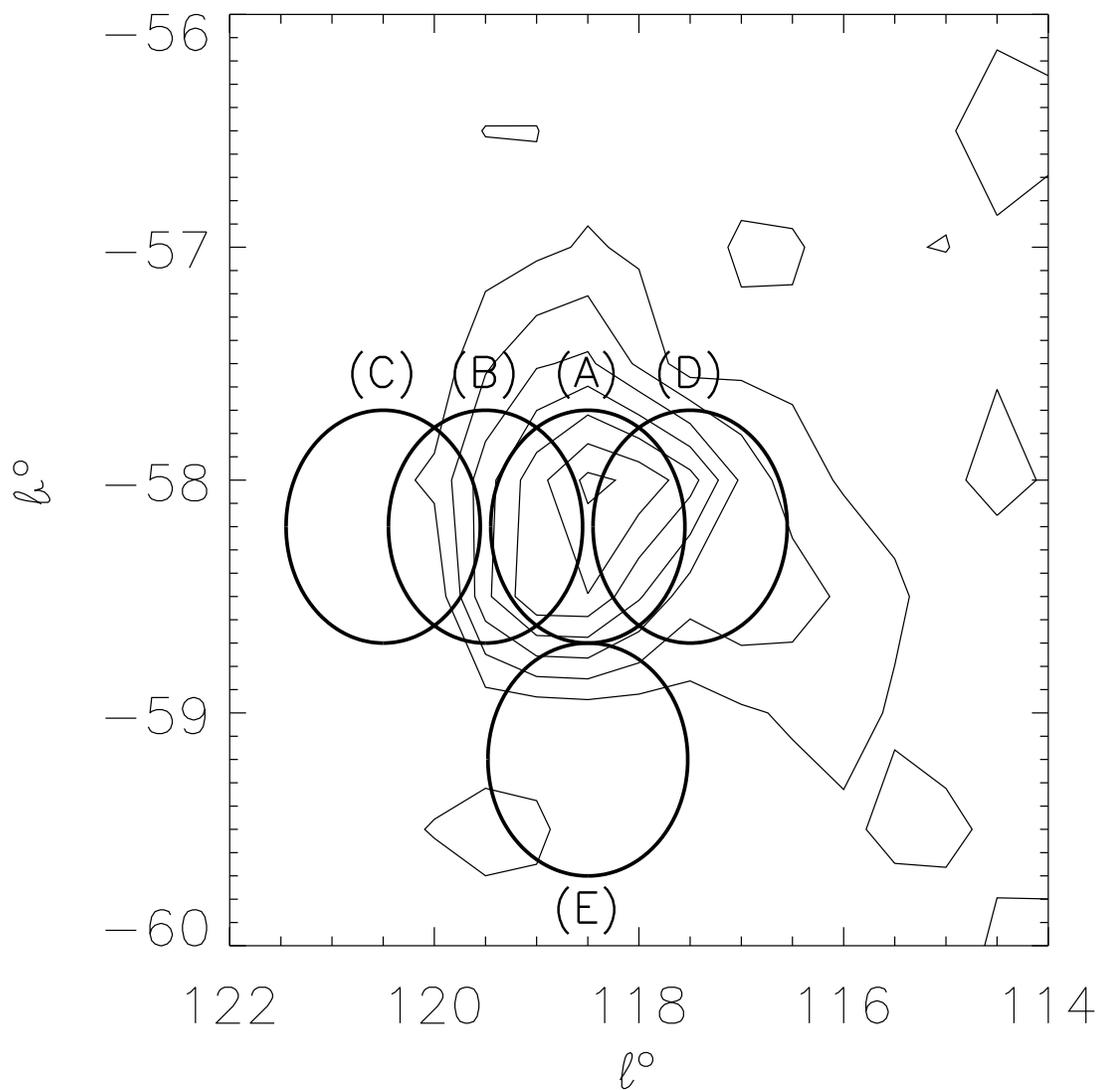}

\caption{WHAM observation beams for HVC 532 overlayed on H~I 21 cm
 contours.  The contours correspond to H~I emission between -450 km
 s$^{-1}$ $<$ V$_{lsr}$ $<$ -250 km s$^{-1}$.}

\end{figure}

\clearpage

\begin{figure}[ht]
\plotone{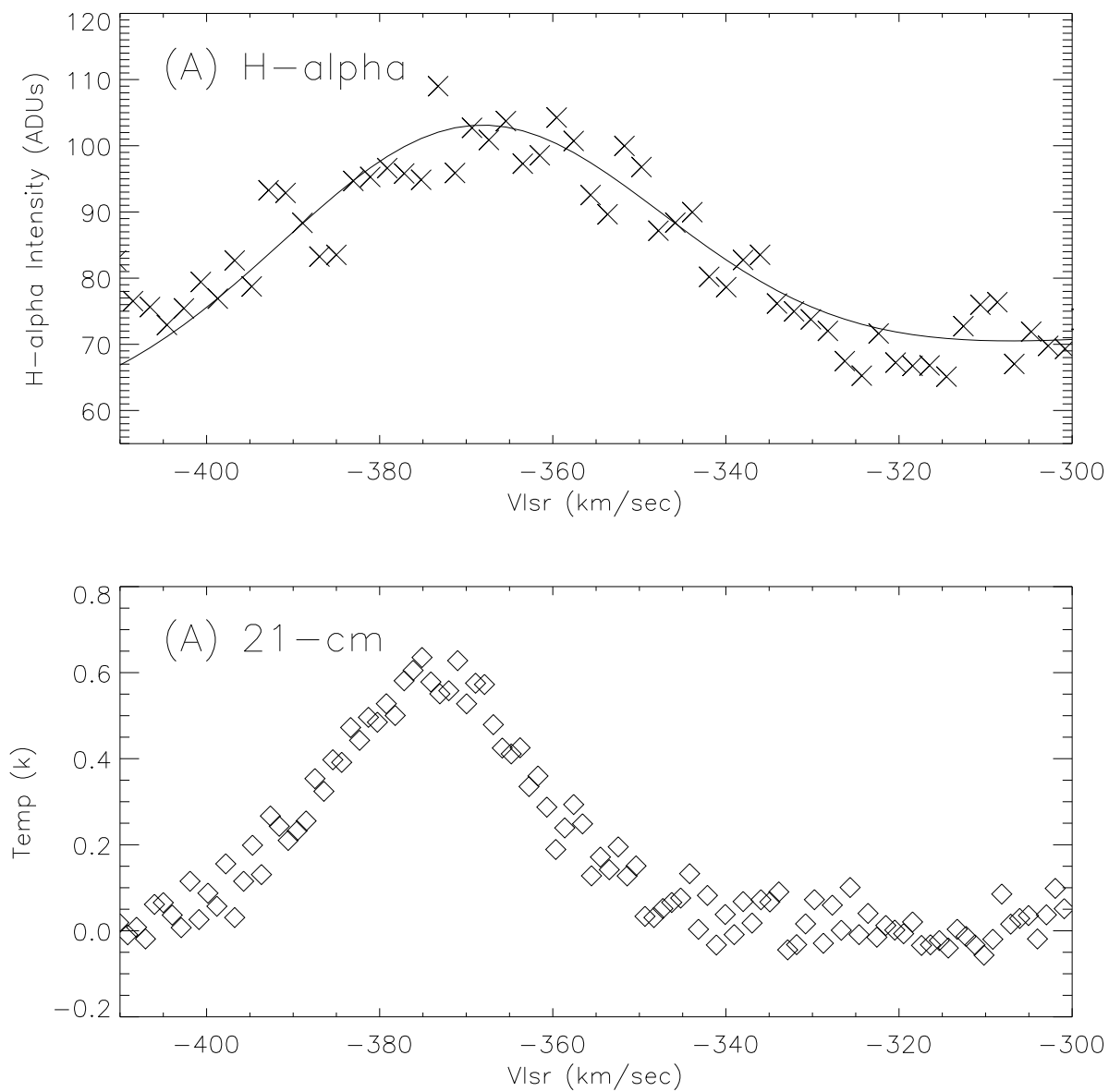}

\caption{The upper panel is the H$\alpha$ spectrum for HVC
	532, (l,b) = (118.5,-58.2), direction (A) from Figure 1.
	The emission feature has an intensity of 0.14 R.  The
	lower panel shows the corresponding 21-cm spectrum from
	LDS. There is a slight velocity offset between the two, and
	the line is wider in H$\alpha$.}

\end{figure}

\clearpage

\begin{figure}[ht]
\plotone{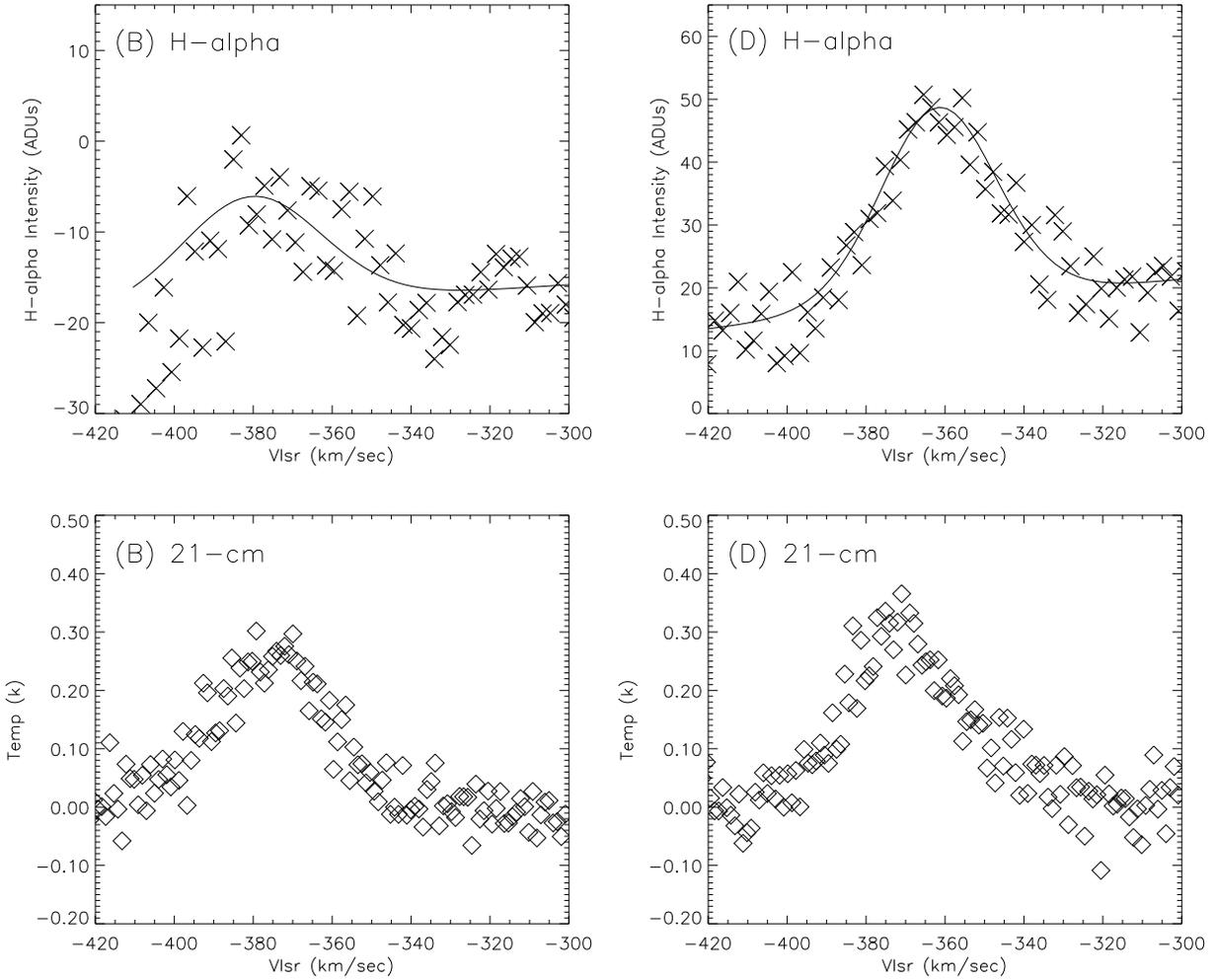}

\caption{The upper left panel shows H$\alpha$ from HVC 532
	(l,b) = (119.5,-58.2), direction (B) in Figure 1.  The lower
	left is the corresponding 21-cm emission from LDS.  The upper right
	panel shows the H$\alpha$ spectrum for HVC 532 (l,b) =
	(117.5,-58.2), direction (D) in Figure 1.  Below this is the
	corresponding 21-cm spectrum.}

\end{figure}

\clearpage

\begin{figure}[ht]
\plotone{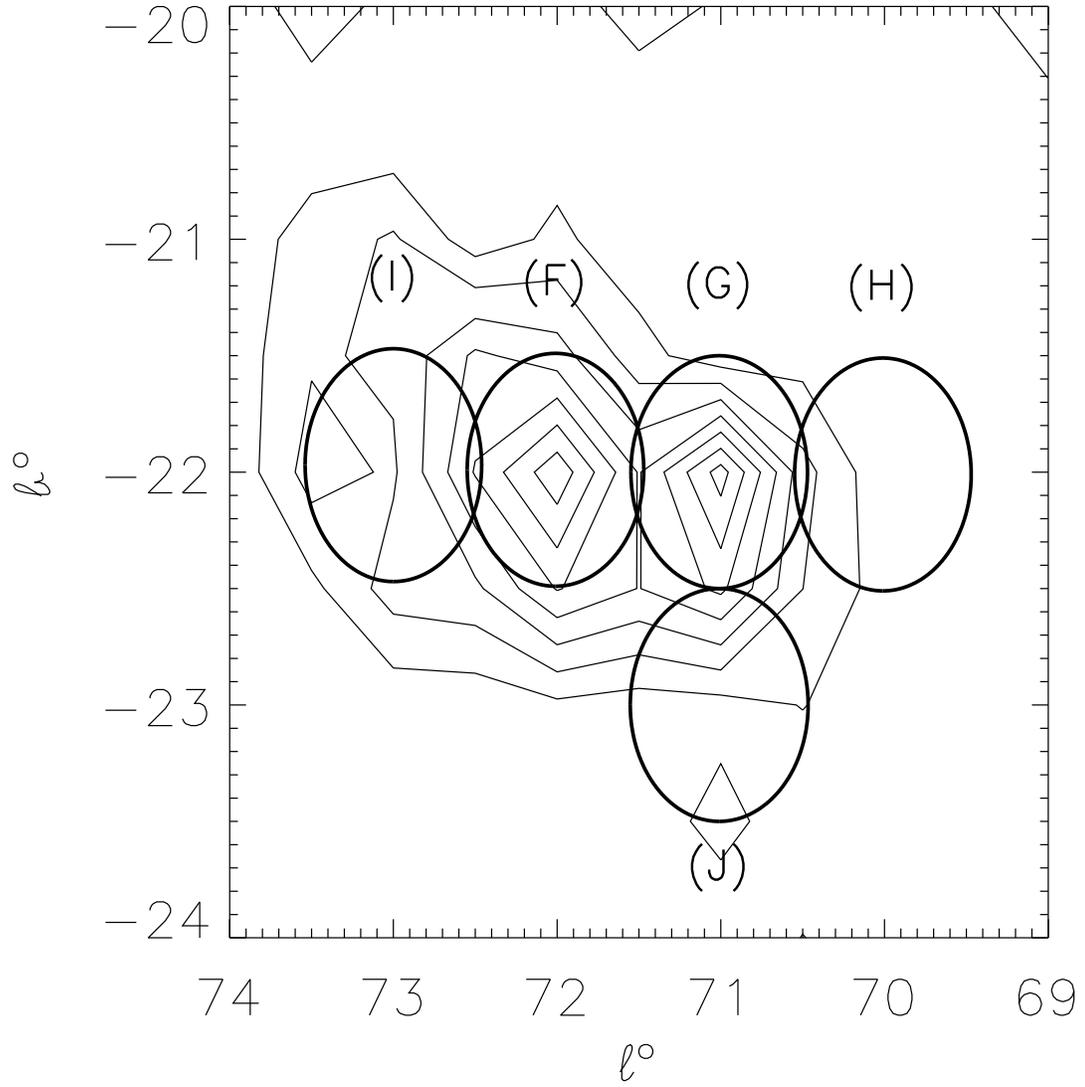}

\caption{WHAM observations of HVC 394 plotted over 21 cm contours 
(-450 km s$^{-1}$ $<$ V$_{lsr}$ $<$ -250 km s$^{-1}$). }

\end{figure}

\clearpage

\begin{figure}[ht]
\plotone{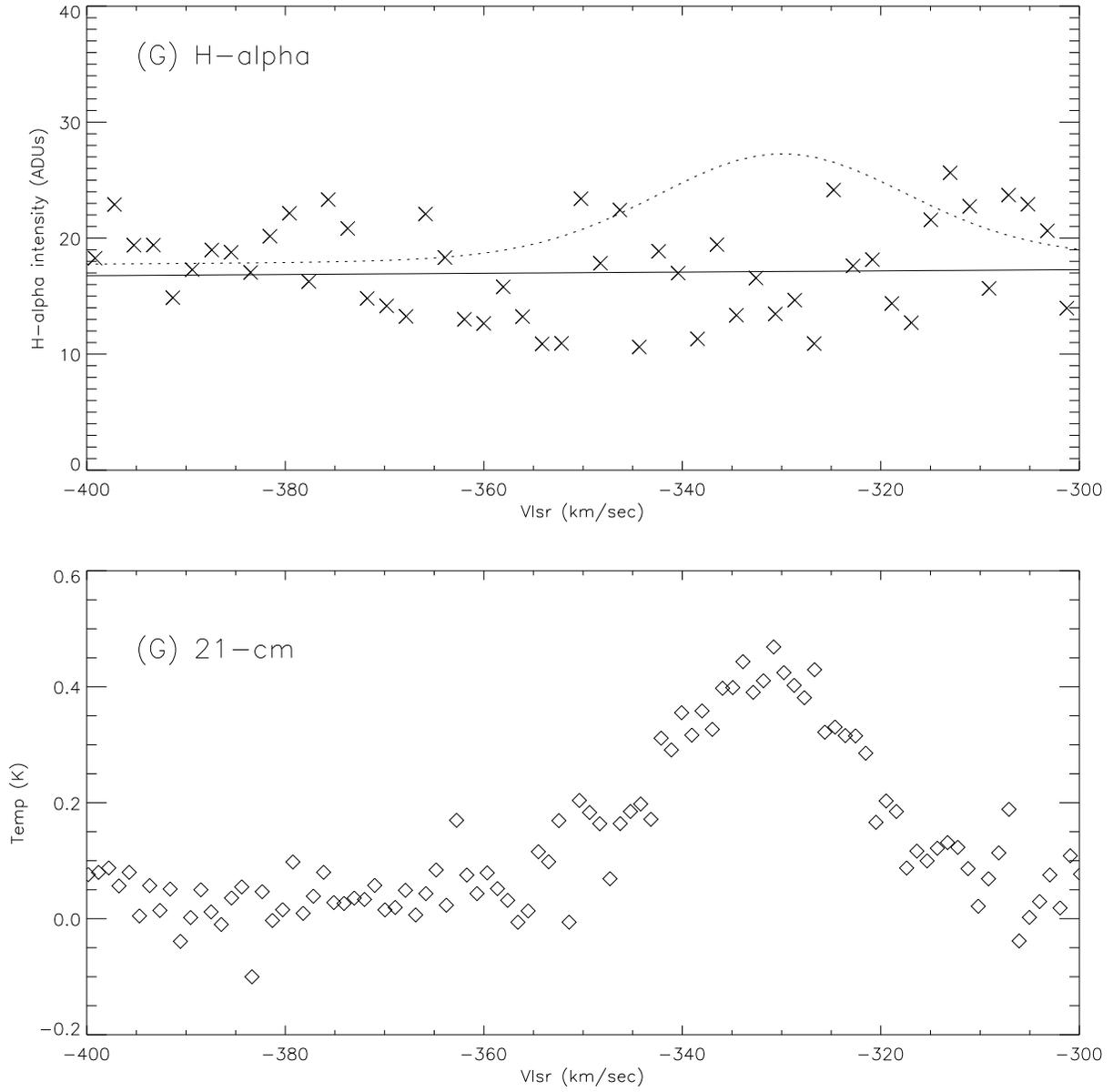}

\caption{The upper panel shows the H$\alpha$ spectrum for HVC
	394 (l,b) = (71.0, -22.0), direction (G) in Figure 4.  The
	solid line is the best fit line, and the dotted line shows
	the upper limit on emission.  The lower panel shows the
	21-cm spectrum for the same direction.}

\end{figure}
\clearpage


\begin{thebibliography}{}


\bibitem[Bland-Hawthorn \& Maloney 1999]{Bla99} Bland-Hawthorn, J.,
\& Maloney, P. R.  1999, \apj, {\em 510}, L33.

\bibitem[Blitz et al. (1999)]{Bli99} Blitz, L., Spergel, D. N., Teuben,
P. J., Hartmann, D., \& Burton, W. B.  1999, \apj, {\em 514}, 818.

\bibitem[Braun & Burton (1999)]{Bra99} Braun, R., \& Burton, W. B. 1999,
A\&A, {\em 341}, 437.

\bibitem[Bregman 1980]{Bre80} Bregman, J. N. 1980, \apj, {\em 236},
577.

\bibitem[Haffner, Reynolds, \& Tufte (1999)]{Haf99} Haffner, L. M.,
Reynolds, R. J., \& Tufte, S. L.  1999, \apj, {\em 523}, 223.

\bibitem[Hartmann \& Burton (1997)]{Har97} Hartmann, D., \& Burton,
W. B.  1997, Atlas of Galactic Neutral Hydrogen (Cambridge: Cambridge
University Press).

\bibitem[Madsen et al. 2001]{Mad01} Madsen, G. J., Reynolds, R. J.,
Haffner, L. M., Tufte, S. L., \& Maloney, P. R.  2001, \apj, {\em 560}, L135.

\bibitem[Oort 1970]{Oor70} Oort, J. H. 1970, A\&A, {\em 7}, 381.

\bibitem[Reynolds et al. (1998)]{Rey98} Reynolds, R. J., Tufte,
S. L., Haffner, L. M., Jaehnig, K., \& Percival, J. W.  1998, PASA, {\em 15},
14.

\bibitem[Shull et al. 1999]{Shu99} Shull, J. M., Roberts, D., Giroux,
M. L., Penton, S. V., \& Fardal, M. A.  1999, \aj, {\em 118}, 1450.

\bibitem[Tufte 1997]{Tuf97} Tufte, S. L., PhD Dissertation, Univ. of
Wisconsin - Madison.

\bibitem[Tufte, Reynolds, \& Haffner (1998)]{Tuf98} Tufte, S. L.,
Reynolds, R. J., \& Haffner, L. M.  1998, \apj, {\em 504}, 773. 

\bibitem[Vogel et al. 1995]{Vog95} Vogel, S. N., Weymann, R., Rauch,
M., \& Hamilton, T.  1995, \apj, {\em 441}, 162.

\bibitem[Wakker, B. P. (2001)]{Wak01} Wakker, B. P.  2001, \apjs, {\em
136}, 463.

\bibitem[Weiner et al. (2001)]{Wei01} Weiner, B. J., Vogel, S. N.,
Williams, T. B.  2001, in Gas \& Galaxy Evolution, ASP Conf. Proc., 
Vol. 240, eds. Hibbard, J. E., Rupen, M., \& van Gorkom,
J. H., 515.

\bibitem[Weymann et al. 2001]{Wey01} Weymann, R. J., Vogel, S. N.,
Veilleux, S., \& Epps, H. W.  2001, {\em 561}, 559.

\end{thebibliography}
\end{document}